Coherent control and high-fidelity readout of chromium ions in commercial silicon carbide

Berk Diler[1], Samuel J. Whiteley[1,2], Christopher P. Anderson[1,2], Gary Wolfowicz[1,3], Marie E. Wesson[2], Edward S. Bielejec[4], F. Joseph Heremans[1,3], David Awschalom[1,2,3,*]

[1]Pritzker School of Molecular Engineering, University of Chicago, Chicago, IL, USA

[2]Department of Physics, University of Chicago, Chicago, IL, USA

[3]Center for Molecular Engineering and Materials Science Division, Argonne National Laboratory, Lemont, IL, USA

[4]Sandia National Laboratories, Albuquerque, NM, USA

**Abstract**

Transition metal ions provide a rich set of optically active defect spins in wide bandgap semiconductors. Chromium ($Cr^{4+}$) in silicon-carbide (SiC) produces a spin-1 ground state with a narrow, spectrally isolated, spin-selective, near-telecom optical interface. However, previous studies were hindered by material quality resulting in limited coherent control. In this work, we implant Cr into commercial 4H-SiC and show optimal defect activation after annealing above 1600 °C. We measure an ensemble optical hole linewidth of 31 MHz, an order of magnitude improvement compared to as-grown samples. An in-depth exploration of optical and spin dynamics reveals efficient spin polarization, coherent control, and readout with high fidelity (79%). We report $T_1$ times greater than 1 s at cryogenic temperatures (15 K) with a $T_2^* = 317$ ns and a $T_2 = 81$ μs, where spin dephasing times are currently limited by spin-spin interactions within the defect ensemble. Our results demonstrate the potential of $Cr^{4+}$ in SiC as an extrinsic, optically active spin qubit.

**Introduction**

Optically active defect spins in wide bandgap semiconductors have attracted much attention as candidate qubits for quantum information[1,2]. These defects hold promise due to their long lived, highly localized ground state spin states. They can be manipulated using a number of quantum control mechanisms[3] and interfaced with optics through atom-like spin-selective transitions for long-distance quantum communication and entanglement[4]. The wide range of available point defects, including complexes and vacancies in diamond[4–6] and silicon-carbide (SiC)[7–10] as well as rare-earth ions in oxides[11], offer many possible electronic, spin, and optical properties depending on their host materials. As each host provides different benefits, it is desirable to decouple the defect properties from their hosts for reliable defect qubit integration into semiconducting devices.

The prototypical host-independent optically active defects are rare-earth ions, since the transitions within their partially filled *f*-orbitals are shielded by their outer electronic shells[12]. However, their magnetic fine structure within the orbitals due to the crystal field is highly host dependent and is difficult to predict[13]. Furthermore, the weak dipole moment of these orbitals result in millisecond long excited state lifetimes which makes optical spin polarization and detecting a single emitter challenging without integrated nanophotonics[14–16].

Transition metal ions have shown to be host-independent by preserving the physics of the defect center across multiple platforms[10]. This is enabled by the similar defect *d*-orbital electronic structures between various hosts that share lattice symmetry and have similar bond strength[17]. Transition metal systems in SiC have already attracted attention as optically active spin qubits such as $V^{4+}$ [18,19] and $Mo^{5+}$ [20]. Both defects possess only one electron in their *d*-orbitals, resulting in a 3d$^1$ and 4d$^1$ configuration, respectively, with a spin-½ ground and excited states. Under the SiC ligand symmetry the ground state is expected to be comprised of multiple close orbitals which could result in short spin-lattice relaxation ($T_1$) times[19,21].

By contrast, chromium defects in SiC in the 4+ charge state (Cr$^{4+}$) are in the 3d$^2$ electronic configuration, forming a $^3A_2$, orbital singlet spin triplet ground state and a $^1E$ spin singlet orbital doublet excited state. The non-degenerate nature of the ground state orbital yields long spin $T_1$ times while the low nuclear spin density of the SiC host should enable long spin coherence times ($T_2$). Even though the optical transition between $^1E$ and $^3A_2$ states is spin forbidden, resulting in a long decay time (146 μs), both states are formed by the different electron configuration of the same orbitals, creating a weak Jahn-Teller coupling[22] which leads to a very high Debye-Waller factor of at least 75%[10]. This well-isolated transition lies in the near-telecom range (1070 nm), making Cr$^{4+}$ particularly interesting for in-fiber long distance quantum communication with integrated spins. However, the previous transition metal ion studies in SiC including Cr$^{4+}$ have been mostly conducted on as-grown samples[10,18,20] and their coherent ground state spin characterization is yet to be demonstrated.

In this work, we study the creation of Cr$^{4+}$ defect spins in commercial SiC through implantation and annealing, a critical step towards accurate three-dimensional localization of spin defects for device integration[23]. Since the optical transition is rather insensitive to environmental perturbations like strain, the fine structure of an ensemble of Cr$^{4+}$ can be resolved under resonant excitation. This property combined with the Λ-like configuration of the spin forbidden optical transitions makes ensemble spin polarization and readout possible. We therefore demonstrate high fidelity coherent control of the Cr$^{4+}$ ensemble by Rabi oscillations with contrast exceeding 79%. We then characterize the ground state spin coherence times of $Cr^{4+}$ in SiC and find long longitudinal spin relaxation times ($T_1$) at cryogenic temperatures and spin-dephasing relaxation time ($T_2$) over 80 μs, similar to other spin defect ensembles in SiC[24]. The transportable nature of the electronic configuration and the lattice insensitive narrow optical transitions combined with long spin coherence times that can be read out with high fidelity, makes $Cr^{4+}$ in SiC and more generally, the $d^2$ configuration in strong tetrahedral ligand field, a promising optically active, transportable and localizable spin qubit.

**Results**

Creation and activation of $Cr^{4+}$ in commercial SiC is a critical step for devices with integrated, optically accessible spins. To this end we introduce isotopically pure, nuclear spin free $^{52}Cr$ atoms into commercially purchased 4H-SiC substrates through implantation at elevated temperatures up to 700 °C. We believe that the implanted chromium atoms sit mainly in interstitial sites of the SiC lattice. During the subsequent annealing, Cr atoms move to the silicon (Si) site and form bonds with the surrounding carbon atoms (Fig. 1a). This generates the $Cr^{4+}$ level structure shown in Fig. 1b (defect activation). The two different Si lattice sites in 4H-SiC result in two distinct $Cr^{4+}$ species: $Cr_A$ (quasi-cubic k-site) and $Cr_C$ (quasi-hexagonal h-site)[22]. To investigate the electrical activation of the defect ensembles, we anneal different samples from the same implanted substrate between 800 and 1800 °C and then measure the zero-phonon line (ZPL) intensities of $Cr_A$ and $Cr_C$ (Fig. 1c). $Cr^{4+}$ activation, not observable when the samples are not annealed, increases monotonically as a function of annealing temperature until 1600 °C where it saturates.

For coherent characterization of the defect, we create $Cr^{4+}$ in a wider range of depth in SiC by combining multiple high energy ion implantations (see Methods) followed by 1800 °C annealing to reduce the overall $Cr^{4+}$ density while maximizing its signal. Off-resonant photoluminescence (PL) at 30 K demonstrates successful activation of $Cr^{4+}$ in this sample (Fig. 1d). The weak spin-forbidden transition of $Cr^{4+}$ under resonant excitation is challenging to detect in the presence of other four orders of magnitude brighter optically active residual defects (such as the divacancy[8]) created during substrate growth and ion implantation. However, the long-lived excited state of $Cr^{4+}$ enables temporal separation of the relevant PL from the background (see SI) by only integrating over the transient signal. We measure the photoluminescence excitation (PLE) spectrum by resonantly exciting $Cr^{4+}$ ZPL and collecting the transient phonon sideband (Fig. 1e) at 30 K. A fit to two Gaussian functions for the $m_s = 0$ and $m_s = \pm 1$ sublevels with a known ground state crystal field splitting ($D$) for $Cr_A$ (1063 MHz) yields an average PLE linewidth of 5.1(4) GHz with the $m_s = 0$ linewidth roughly two times larger than the $m_s = \pm 1$ linewidth. The origin of this behavior is unknown and may be a result of mass related shift due to $^{13}C$, $^{29}Si$ and $^{30}Si$ isotopes[19,25,26]. We observe a similar trend for $Cr_C$ (see SI) where the individual PLE lines are more resolved due to a larger $D$. The overall sharper inhomogeneous linewidth of the implanted sample compared to the as-grown sample[10] allows for better spin sublevel resolution and is likely due to an improved spatial strain profile.

We then characterize the ground state spins of $Cr^{4+}$ in SiC through resonant hole burning and hole recovery at 15 K where the spin $T_1$ is longer than the optical lifetime. The population of a spin sub-ensemble excited by the narrow line resonant laser will be redistributed among the other ground state spin sublevels after spontaneous emission, leading to hole burning and spin polarization. Since the polarized sub-population is off-resonant relative to the probe laser, the defect goes dark resulting in a spin selective optical contrast. We can recover the hole by reintroducing the polarized population into the optical cycle in two ways, either by using a second laser color to probe the polarized sublevels or by driving ground state spins with microwaves. A second laser tone generated by a phase modulator recovers the $Cr_A$ hole at a $D$ of 1062.7(4) MHz (Fig. 2a). A Lorentzian fit to the data extracts a 31(2) MHz ensemble hole linewidth which is more than an order of magnitude narrower than the as grown samples[10], demonstrating the superior material quality of the implanted and annealed sample. However, the sharp hole linewidth is also two orders of magnitude narrower than the inhomogeneous optical linewidth, indicating only 1% of the created $Cr_A$ defects are accessed under resonant excitation. One could increase the net ensemble polarization by further reducing the inhomogeneous linewidth through material improvements. The hole can also be recovered by microwave mixing of the ground state spin sublevels, enabling optical detection of magnetic resonance (ODMR) (Fig. 2b). Stray magnetic fields Zeeman split the $m_s = \pm1$ ODMR by ~ 5 MHz centered around $D$ = 1063.11(1) MHz (Fig. 2b). A Lorentzian fit to the non-power broadened ODMR reveals a linewidth of 1.32(2) MHz. Because the 31 MHz optical hole linewidth is not confined by the ODMR linewidth, the optical coherence is likely excited state limited. The 4 orders of magnitude difference between the measured and lifetime limited hole linewidth (~2kHz) may be a result of decoherence induced by the degenerate excited state orbital doublet structure or spectral diffusion due to charge fluctuations[27]; the exact mechanism would require further investigation.

A *c*-axis magnetic field lifts the optical degeneracy of the $m_s = \pm1$ sublevels by Zeeman splitting while the S = 0 excited state remains unaffected (Fig. 1b). It is worth noting that the population can become trapped in the unprobed third level after using either the optical sideband probe or the microwave mixing, as these techniques can only address one sub-level at a time. By applying both an optical sideband probe and a microwave tone together, all states can be addressed simultaneously resulting in hole recovery (Fig. 2c). Using a second laser to trap the population into the third sublevel provides a path for efficient spin polarization under a magnetic field.

For coherent control and readout of $Cr^{4+}$, we polarize a subpopulation into the $m_s$ = +1 sublevel by tuning the primary laser color to be resonant with $m_s$ = 0 state and adding a sideband tone to optically excite the $m_s$ = -1 sublevel (see SI). Once polarized, we coherently control the subpopulation using resonant microwave rotations within the $m_s$ = 0, +1 manifold and read out the $m_s$ = 0 population with a short (50 µs) primary laser excitation (Fig. 3a). We measure Rabi rotations at $B$ = 158 G by varying the microwave excitation length and obtain a contrast of 63(1)% with an envelope decay time of 4.76(7) µs (Fig. 3b).

To characterize the ground state spin coherence times, we perform Ramsey interferometry and Hahn echo measurements. For the Ramsey measurement, we detune the resonant microwave frequency by 5 MHz and delay the time between two $\pi_x/2$ pulses. The characteristic oscillations at the detuning frequency have an envelope with a decay time of $T_2^*$ = 307(17) ns (Fig. 3c), in agreement with the ODMR linewidth measurements in Fig. 2b. We measure the $T_2$ coherence time by ($\pm \pi_x/2$ - $t_{free}/2$ - $\pi_y$ - $t_{free}/2$ - $\pi_x/2$) echo-pulse sequence. We fit the data to an electron spin echo envelope modulation function where nuclear Larmor precessions are enveloped by a decay function of $e^{-(\frac{t}{T_2})^n} \prod(1 - K_a \sin^2(\pi \omega_a \tau))$, with the precession amplitudes ($K_a$), frequencies ($\omega_a$) and decay power ($n$) and time ($T_2$) as free fit parameters[28]. We recover the characteristic oscillations at the current $B$ field for $^{13}$C (87.5(3) kHz) and $^{29}$Si (68.0(1) kHz) as well as $T_2$ = 81(2) μs with a decay power dependence of $n$ = 1.9(1) (Fig. 3d). The average defect density in our sample is roughly 3 x 10$^{16}$ atoms/cm$^3$ (see SI) and the measured $T_2$ time is in line with similarly dense divacancy ensembles in SiC[24]. This indicates that the observed coherence times are likely limited by electron spin-spin interactions between defects, and further reduction of the Cr$^{4+}$ and background defect densities could result in millisecond long coherence times[28].

Finally, we report that by sacrificing overall signal and reducing the probe time from 50 μs to 1 μs we increase the Rabi contrast to 79(2)% contrast (Fig. 3f). This places a lower bound on the ensemble polarization of at least 77%. The contrast, however, is lower than our estimate (see SI) and we believe a spin-reset mechanism induced by the probe itself could be responsible, therefore, further exploration of this effect could increase the readout fidelity.

The spin characterization measurements reported above are enabled by the careful study of the rates responsible for the Cr$^{4+}$ population dynamics. One such rate is the excited state lifetime ($T_{opt}$) under resonant excitation which we measure at 30 K by histogramming the transient PLE to find $T_{opt}$ = 156.3(5) μs (Fig. 4a). This is in agreement with the off-resonant excited state lifetime[10] (see SI). Based on this measurement, our integration window for the collected photons in the transient readout is 155 μs long.

To understand the ground state spin dynamics, we measure the spin $T_1$ time at $B$ = 0 G by population inversion relaxation using a microwave π-pulse. By repeating this measurement for temperatures between 15 – 30 K and fitting each data set to an exponential decay function, we extract $T_1$ times and use this dataset to distinguish between Orbach and Raman processes[29] (Fig.

4b). For the Orbach process, we use $\frac{1}{T_1} = A\, e^{-\frac{E}{k_B T}}$ as our fit function where A and E, the energy difference to a low-lying excited state, are fit parameters. For the Raman process we use $\frac{1}{T_1} = A\,(T - \Delta T)^n$ where A and ΔT, a constant temperature offset of the sample, are fit parameters for various fixed odd integers n. The Orbach model fits the data poorly and returns an energy gap of E = 20(1) meV which is a factor of 50 lower from the first excited state energy and a factor of 4 lower than the closest phonon line energy[22], indicating that Orbach process is not the likely explanation for $T_1$ decay mechanism. Out of the Raman processes, the most likely explanation is the n = 9 process (see SI), indicating that the $T_1$ is spin-orbit interaction limited. The longest measured spin $T_1$ time is 1.6(3) s long at T = 15 K and offers a very long window to both initialize and coherently manipulate the spin.

Finally, we time resolve the polarization dynamics of $Cr^{4+}$ to investigate defect initialization. We sweep the resonant excitation time using the same parameters in Fig. 3b-d and measure the populations of the $m_s$ = 0 and the $m_s$ = +1 spin sublevels (Fig. 4c). When the resonant excitation time of the $m_s$ = 0 is less than the optical lifetime of 155 µs, the population in the excited state transferred from the $m_s$ = 0 increases. Past that point in time, the $m_s$ = 0 population depletes as the sub-ensemble polarizes into the $m_s$ = +1. Using a π-pulse to coherently swap the populations between the $m_s$ = 0 and the $m_s$ = +1 states, we measure the polarized population within the $m_s$ = +1 sublevel which increases monotonically with the laser excitation time. The difference of the two traces is the optical contrast shown in Fig. 4d. An exponential fit to the data yields a rise time of 1.27(3) ms, corresponding to polarization within ~ 10 optical lifetimes with maximum contrast at 64(2)%. In order to achieve the reported 63% contrast for a 50 µs probe time, we polarized the defect for 5 ms. Reducing the optical lifetime through Purcell enhancement in an optical cavity would significantly decrease this initialization time.

**Discussion**

In summary, we have demonstrated successful creation and activation of $Cr^{4+}$ ions with exceptional optical and spin properties in commercial semi-insulating SiC substrates. This method could be used to introduce other transition metal ions such as V[18,19], Mo[20] and possibly Cu into SiC and GaN[10]. $Cu^{4+}$ in a tetrahedral lattice with a strong ligand field as found in SiC, would be in the $3d^7$ electronic configuration, and could produce the electronic structure of highly studied $Al_2O_3$:$Cr^{3+}$ (Ruby)[30–32].

The reported defect creation technique can also be extended to three-dimensional localization of single transition metal ions in SiC for device integration through nano-implantation[33,34]. For example, a single $Cr^{4+}$ can be detected by placing it into a photonic cavity to reduce the excited state lifetime by Purcell enhancement[15]. Within such a device, the already large fraction of indistinguishable photons in the near-telecom ZPL would be further enhanced[4]. A dense ensemble of $Cr^{4+}$ ions within a cavity may also achieve population inversion through the second-excited state to realize a laser[35]. All these possible photonics applications readily come with integrated spins that can be initialized and read out with a high fidelity of at least 79%. Furthermore, the measured ensemble spin coherence time $T_2$ of 81 μs is spin-spin limited, showing that the potential of an isolated $Cr^{4+}$ in SiC as spin qubits[28].

Finally, our results can be generalized to a broader set of transition metal ions in $d^2$ electronic configuration in a host that provides a strong tetrahedral ligand field. The 31 MHz ensemble optical linewidth and the 79% readout fidelity combined with $T_1$ time greater than 1 s makes these class of transition metal electronic structures a promising set of transportable optically active spin qubits.

**Methods**

**Sample Preparation**

In Fig. 1c the substrates were high purity semi-insulating, on-axis 4H-SiC sourced by Norstel AB. The SiC samples were co-implanted with $^{52}$Cr at 190 keV energy with a dose of 5x10$^{11}$ cm$^{-2}$, and $^{12}$C at 100 keV energy with a dose of 5x10$^{11}$ cm$^{-2}$ by CuttingEdge Ions, LLC. Both ions were hot implanted with a 7° tilt at a temperature of 600 °C. Chromium atoms are calculated to have an average depth of 110 nm and a 30 nm straggle using the program Stopping Range of Ions in Matter (SRIM). All samples were subsequently cleaned with organic solvents. A photoresist layer (AZ1518) is spun ~2 µm thick, baked at 95 °C for 1 minute then 350 °C for 30 minutes, as a protective surface coating by forming a graphite cap during the high temperature annealing. The samples are annealed in a tube furnace with high purity Ar at 800-1400 °C for 15 minutes with ramp rates of 100 °C per hour. The annealing of the samples at 1500-1800 °C was performed by Fraunhofer IISB in an Ar environment for 15 minutes with ramp rates of 900 °C per hour. The graphite cap was removed by heating the sample to 750 °C for 30 minutes in air.

The box profile implanted sample in Fig. 1b-c,2-4 used an epitaxial, undoped (intrinsic), 20 µm thick 4H-SiC layer that was grown on a 4° off-axis semi-insulating SiC substrate sourced by Norstel AB. The 4H-SiC sample was implanted with $^{52}$Cr ions in four steps with energies of 10, 5, 2, 1 MeV and doses of (3, 3, 2, 1.5) x 10$^{12}$ cm$^{-2}$, respectively, at a sample temperature of approximately 690 °C at Sandia National Laboratories. After $^{52}$Cr implantation the sample was cleaned, photoresist capped, and annealed at 1800 °C for 15 minutes using the same procedures detailed above.

**Optical and spin characterization experiments**

A closed cycle cryostat (Montana Instruments Nanoscale Workstation), with microwave and optical access is used to cool the sample. A NIR 50x objective (Olympus) is mounted to a heated and isolated shroud within the cryostat. Three in-cryo piezo stages (Attocube), each with a 4 mm range enabled switching between samples and locating microwave striplines.

A Thorlabs TO can 730 nm laser is used for off-resonant excitation. Resonant excitation is achieved by a fiber-coupled, external-cavity diode laser with a tuning range of 1035 –1075 nm and a linewidth < 200 kHz integrated over 50 ms with 18 mW maximum output power (Newfocus Velocity 6700). An electro-optical modulator (iXBlue) with a 12 GHz bandwidth that generates optical side-bands is used for the experiments that needed two colors. A fiber coupled acousto-optical modulator (Aaoptoelectronic) is used to rapidly turn the laser on and off, a second one is added for the $T_1$ data in Fig. 4b to improve the extinction ratio over long wait times.

A vector signal generator (Stanford Research Instruments SG396) with a 30W 0.6 - 2.7 GHz amplifier (Minicircuits) is used for microwave excitation. It is fed to a shorted coplanar wave guide patterned on a custom-made printed circuit board situated behind the sample within the cryostat.

For spectrally resolved measurements a liquid nitrogen cooled InGaAs NIR camera with a spectrometer (Princeton Instruments) is used. For resonant measurements, a superconducting nanowire single photon detector that is optimized for 1.0 - 1.2 μm regime coupled to a 25 μm multimode fiber (Quantum-Opus LLC) is used. The average dark count resulting from thermal photons slowly varying between 6 – 9 k counts/s were measured before an experiment and subtracted from the data. The detection signal is gated using a microwave switch (Minicurcuits) and counted using a data acquisition system (National Instruments).

An arbitrary waveform generator (Swabian Instruments) is used to synchronize timings of the laser excitation, probe generation, microwave excitation with IQ control, collection and counter gating.

## Acknowledgements

The authors thank S. Bayliss, M. Fataftah, D.W. Laorenza, M.K. Wojnar, T. Fidler and A. Bauer for useful discussion. This project was supported by the US Department of Energy, Office of Science, Basic Energy Sciences, Materials Sciences and Engineering Division. This work was performed, in part, at the Center for Integrated Nanotechnologies, an Office of Science User Facility operated for the U.S. Department of Energy (DOE) Office of Science. Sandia National Laboratories is a multimission laboratory managed and operated by National Technology & Engineering Solutions of Sandia, LLC, a wholly owned subsidiary of Honeywell International, Inc., for the U.S. DOE's National Nuclear Security Administration under contract DE-NA-0003525. The views expressed in the article do not necessarily represent the views of the U.S. DOE or the United States Government.

## Author Contributions

**Competing Interests**

The authors declare that they have no competing financial interests.

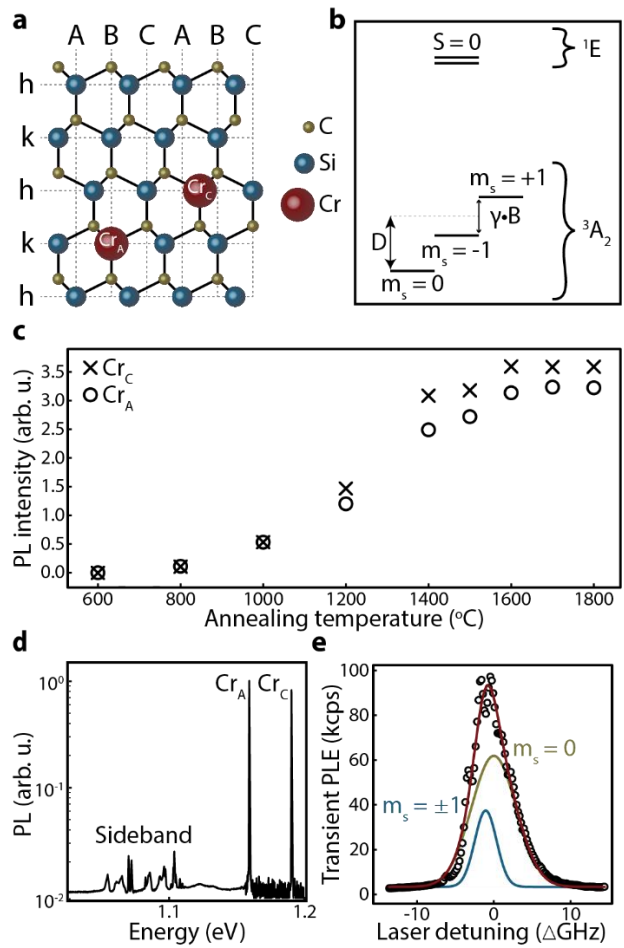

**Fig. 1** Structure, creation and spectroscopy of chromium defects in 4H-SiC **a**, An illustration of substitutional $Cr^{4+}$ ions in silicon sites of a 4H-SiC lattice adapted from Ref [36]. **b**, The electronic level structure of $Cr^{4+}$ in SiC. The $m_s = 0$ and $m_s = \pm1$ sublevels are split by the crystal field ($D$), and under a magnetic field, the $m_s = \pm1$ sublevels are Zeeman split. **c**, $Cr^{4+}$ activation as a function of annealing temperature is measured by integrating the zero-phonon lines (ZPLs) intensity of the photoluminescence under off-resonant (730 nm) excitation at $T = 30$ K. **d**, A photoluminescence spectrum of the sample used for spin and optical control at $T = 30$ K. $Cr_A$ and $Cr_C$, two different sites of $Cr^{4+}$ in 4H-SiC, ZPL's and their sidebands can be observed. **e**, $Cr_A$ photoluminescence excitation (PLE) at $T = 30$ K is measured by sweeping the resonant laser and recording the transient sideband signal in counts per second (cps). The PLE is fit to two Gaussian peaks with a known $D$ = 1063 MHz splitting[10]. The full width at half maxima are 6.87(27) GHz for $m_s$ = 0 and 3.34(39) GHz for $m_s$ = ±1 peak. The one sigma errors of the data are smaller than the point size and are not displayed.

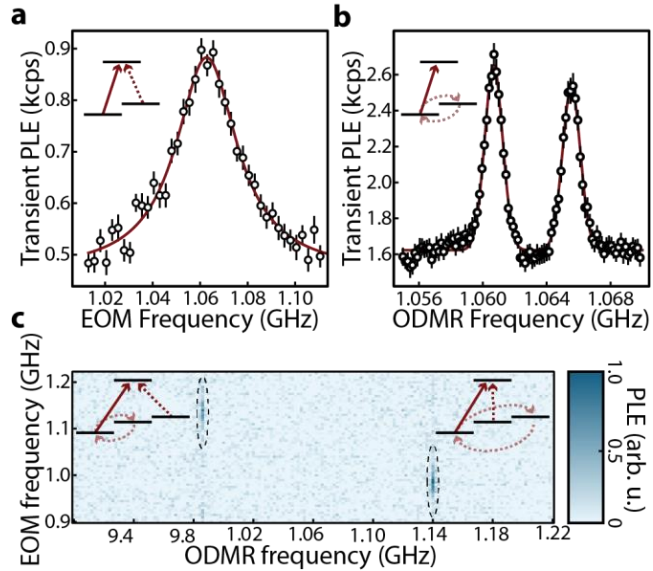

**Fig. 2** Hole burning and recovery of 4H-SiC:$Cr_A$ at $T$ = 15 K. At $B$ = 0 G primary resonant laser color burns a hole, polarizing a subpopulation into the unprobed level, dimming the PLE; shown with a continuous arrow in the level diagram. **a**, Hole recovery of 4H-SiC:$Cr_A$. After hole burning, a second laser tone is scanned, shown with the dashed line in the level diagram. When it is resonant with the unprobed level, the hole is recovered. Lorentzian fit to the recovery is centered at $D$ = 1062.7(0.4) MHz with a full width at half maximum of 31(2) MHz. The laser power is reduced to prevent optical power broadening. **b**, ODMR of 4H-SiC:$Cr_A$ at $B \approx$ 1.7 G. A microwave frequency is swept shown with a dashed line in the level diagram and when it is resonant with the unprobed level PLE is recovered by spin mixing. Stray magnetic fields split the $m_s$ = +1 and $m_s$ = -1 by ~ 5 MHz centered around $D$ = 1063.11(1) MHz. ODMR peaks have a full width at half maximum of 1.32(2) MHz. **c**, Simultaneous optical and magnetic hole recovery at $c$-axis $B$ = 27.5 G where $m_s$ = ±1 are split by 78 MHz. Contrast is achieved when all three ground state sublevels are probed simultaneously with the high contrast area highlighted within dashed ellipses. The error bars are the one sigma of the data.

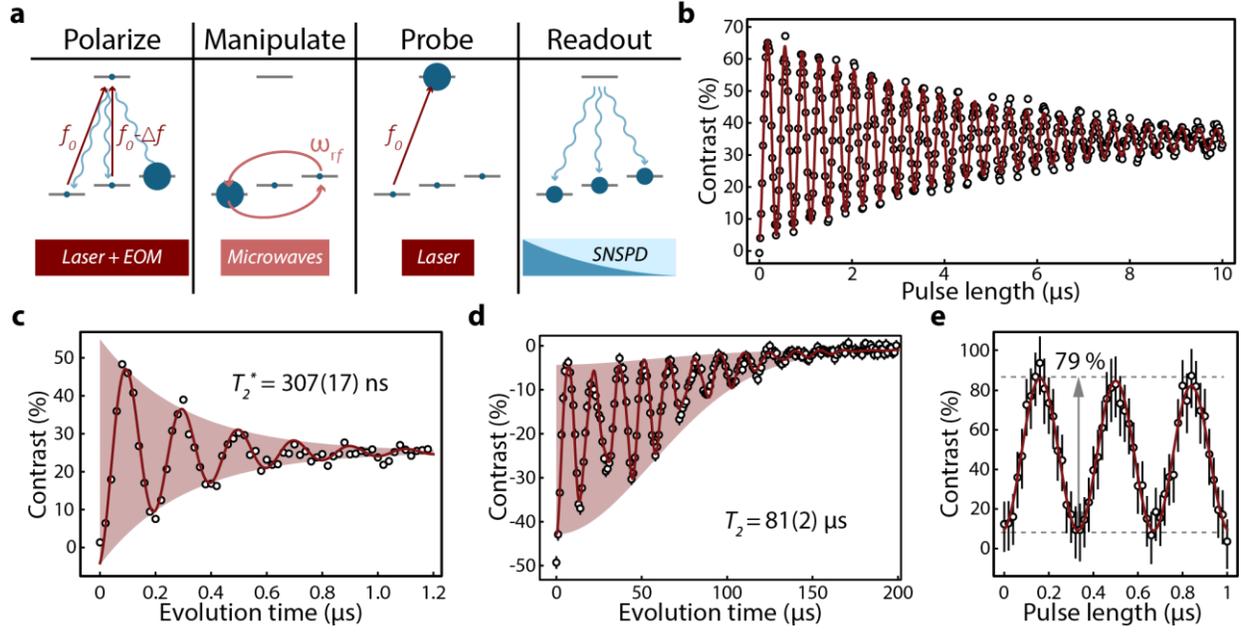

**Fig. 3** Coherent Control of 4H-SiC:Cr$_A$ at $T$ = 15 K and $B$ = 158 G. **a**, An illustration of the coherent control measurement scheme. EOM frequency is tuned to the energy difference between the $m_s$ = 0 and $m_s$ = -1 and a 5 ms selective excitation polarizes the defect into $m_s$ = +1 sublevel. At this point, any microwave pulse sequence can be used to coherently control the population in the $m_s$ = 0, +1 manifold. The population in the $m_s$ = 0 is probed by a short 50 μs laser pulse and the PLE is measured by integrating the transient photoluminescence excitation for 155 μs. **b**, Resonantly driven Rabi oscillations of the ground state spin measured with a 63(1)% contrast and a decay time of 4.76(7) μs. **c**, 5 MHz detuned Ramsey interferometry measurement. A fit to the data reveals a $T_2^*$ = 307(17) ns. **d**, A Hahn echo measurement reveals a $T_2$ = 81(2) μs with characteristic oscillations of $^{13}$C (87.5(3) kHz) and $^{29}$Si (68.0(1) kHz) Larmor precession frequencies at the field. **e**, High contrast Rabi oscillations measured by reducing the probe time to 1 μs. The total signal is reduced but the contrast is increased to 79(2)%. All fits are shown in red and the error bars are the one sigma of the data.

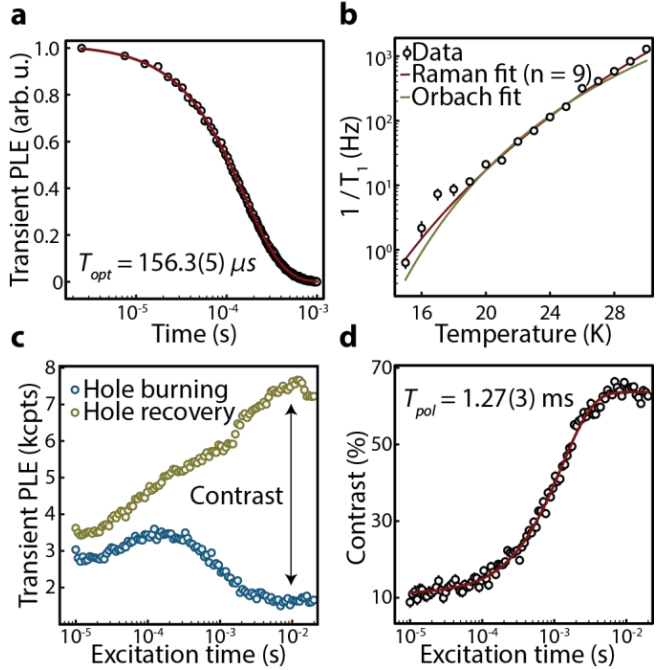

**Fig. 4** Governing rates of 4H-SiC:$Cr_A$. **a**, Normalized photoluminescence excitation decay time at $T$ = 30 K yields a $T_{opt}$ = 156.3(5) µs. **b**, Ground state spin $T_1$. After polarization, the contrast between a $\pi$-pulse and a $2\pi$-pulse is measured at various temperatures. The resulting time-dependences are fit to an exponential decay, where the displayed error bars are the standard errors of the fits. The extracted $T_1$ times are fit to both an exponential (Orbach) and a power (Raman) model. The exponential fit (green line) reveals an E = 20(1) meV with a reduced $\chi^2$ = 4.18. The 9$^{th}$ power Raman fit (red line) reveals a constant temperature miscalibration of $\Delta T$ = 3.2(5) K with a reduced $\chi^2$ = 1.91. **c**, Polarization dynamics at $T$ = 15 K and $B$ = 158 G. $m_s$ = 0 state is probed as a function of polarization pulse length. In the absence of a $\pi$-pulse (blue trace) the depletion of the $m_s$ = 0 state is measured. When a $\pi$-pulse is applied, the accumulation of the polarized population in the $m_s$ = +1 is measured by coherently transferring the population to $m_s$ = 0 (green trace). **d**, The contrast is the difference between the two traces in (c). An exponential fit to the contrast shown in red yields a polarization time of 1.27(3) ms with a maximum contrast of 64(2)%

for a 50 μs probe time. The error bars in plots (a), (c) and (d) are smaller than the mark size and are not shown.